\documentclass[journal,10pt,twocolumn,twoside]{IEEEtran}
\usepackage{times,amsmath,amssymb}
\usepackage{slashbox}
\usepackage{graphicx}
\usepackage{xcolor}
\usepackage{algorithm2e}
\usepackage{algorithmicx}
\usepackage{setspace}
\usepackage[export]{adjustbox}
\usepackage{multirow}
\usepackage[noend]{algpseudocode}

\let\oldin\in

\def\in{\mathrel{
   \mathchoice{\raise.2ex\hbox{$\scriptstyle\oldin$}}
   {\raise.2ex\hbox{$\scriptstyle\oldin$}}
   {\raise.2ex\hbox{$\scriptscriptstyle\oldin$}}
   {\raise.2ex\hbox{$\scriptscriptstyle\oldin$}}}}

\begin{document}

\title{Temporal Attention augmented Bilinear Network for Financial Time-Series Data Analysis}
\author{\IEEEauthorblockN{Dat Thanh Tran\IEEEauthorrefmark{1}, Alexandros Iosifidis\IEEEauthorrefmark{3}, Juho Kanniainen\IEEEauthorrefmark{2} and Moncef Gabbouj\IEEEauthorrefmark{1}} \\
	\IEEEauthorblockA{\IEEEauthorrefmark{1}Laboratory of Signal Processing, Tampere University of Technology, Tampere, Finland\\
        \IEEEauthorrefmark{3}Department of Engineering, Electrical \& Computer Engineering, Aarhus University, Aarhus, Denmark\\
		\IEEEauthorrefmark{2}Laboratory of Industrial and Information Management, Tampere University of Technology, Tampere, Finland\\
		Email:\{dat.tranthanh,juho.kanniainen,moncef.gabbouj\}@tut.fi, alexandros.iosifidis@eng.au.dk}\\
	
}

\maketitle

\begin{abstract}
Financial time-series forecasting has long been a challenging problem because of the inherently noisy and stochastic nature of the market. In the High-Frequency Trading (HFT), forecasting for trading purposes is even a more challenging task since an automated inference system is required to be both accurate and fast. In this paper, we propose a neural network layer architecture that incorporates the idea of bilinear projection as well as an attention mechanism that enables the layer to detect and focus on crucial temporal information. The resulting network is highly interpretable, given its ability to highlight the importance and contribution of each temporal instance, thus allowing further analysis on the time instances of interest. Our experiments in a large-scale Limit Order Book (LOB) dataset show that a two-hidden-layer network utilizing our proposed layer outperforms by a large margin all existing state-of-the-art results coming from much deeper architectures while requiring far fewer computations.

\end{abstract}

\section{Introduction}\label{S:Intro}

Time-series classification and prediction have been extensively studied in different domains. Representative examples include natural language processing \cite{graves2013speech,cho2014learning,bahdanau2014neural}, medical data analysis \cite{lipton2015learning, zabihi2016heart}, human action/behavior analysis \cite{iosifidis2015classspecific,iosifidis2012tnnls, iosifidis2012tifs}, meteorology \cite{xingjian2015convolutional}, finance and econometrics \cite{kaastra1996designing, azoff1994neural, cao2003support} and generic time-series classification \cite{iosifidis2013multidimensional,taieb2016bias}. Due to the complex dynamics of financial markets, the observed data is highly non-stationary and noisy, representing a limited perspective of the underlying generating process. This makes financial time-series forecasting one of the most difficult tasks among time-series predictions \cite{wang2012novel}. The development of both software and hardware infrastructure has enabled the extensive collection of trade data, which is both an opportunity and a challenge for the traders, especially high-frequency traders. Apart from long-term investment, HFT is characterized by high speed and short-term investment horizon. The ability to efficiently process and analyze large chunks of data within relatively short time is thus critical in HFT.

During the past decades, several mathematical models have been proposed to extract financial features from the noisy, nonstationary financial time-series. Stochastic features and market indicators \cite{neely2014forecasting, murphy1999technical} have been widely studied in quantitative analysis.
Notable works include autoregressive (AR) \cite{walker1931periodicity} and moving average (MA) \cite{slutzky1937summation} features, which were later combined as a general framework called autoregressive moving average (ARMA). Its generalization, also known as autoregressive integrated moving average (ARIMA) \cite{box1974some,tiao1981modeling} which incorporates the differencing step to eliminate nonstationarity, is among popular methods for time-series analysis in econometrics. To ensure tractability, these models are often formulated under many assumptions of the underlying data distribution, leading to poor generalization to future observations \cite{poterba1988mean}. In recent years, the development of machine learning techniques, such as support vector regression \cite{sapankevych2009time,cao2003support} and random forest \cite{creamer2004predicting}, have been applied to time-series forecasting problems to alleviate the dependence on such strong assumptions. As a result, these statistical machines often outperform the traditional ARIMA model in a variety of scenarios \cite{kane2014comparison}.

Although the aforementioned machine learning models perform reasonably well, they are not particularly designed to capture the temporal information within time-series data. A class of neural network architecture called Recurrent Neural Networks (RNN) is specifically designed to extract temporal information from raw sequential data. Although RNNs were developed more than two decades ago \cite{hopfield1987neural}, they started to become popular in many different application domains \cite{hinton2012deep,graves2013speech,yue2015beyond} only recently thanks to the development in optimization techniques and computation hardware, as well as the availability of large-scale datasets. Special types of RNNs, such as Long Short Term Memory (LSTM) \cite{hochreiter1997long} and Gated Recurrent Unit (GRU) \cite{chung2014empirical} networks, which were proposed to avoid the gradient vanishing problem in very deep RNN, have become the state-of-the-art in a variety of sequential data prediction problems \cite{hinton2012deep,graves2013speech,yue2015beyond,greff2017lstm}. The beauty of deep neural networks lies in the fact that these architectures allow end-to-end training, which works directly on the raw data representations instead of hand-crafted features. As a result, suitable data-dependent features are automatically extracted, improving the performance and robustness of the whole system.

While deep neural networks in general and LSTM networks, in particular, are biologically inspired and work well in practice, the learned structures are generally difficult to interpret. It has been long known that there exists a visual attention mechanism in human cortex \cite{rensink2000dynamic,corbetta2002control,ungerleider2000mechanisms} in which visual stimuli from multiple objects compete for neural representation. To further imitate the human learning system, attention mechanisms were developed for several existing neural network architectures to determine the importance of different parts of the input during the learning process \cite{xu2015show,bahdanau2014neural,mnih2014recurrent,chen2015abc}. This not only improves the performance of the network being applied to, but also contributes to the interpretation of the obtained result by focusing at the specific parts of the input. For example, in image captioning task \cite{xu2015show}, by incorporating visual attention into a convolutional LSTM architecture, the model explicitly exhibits the correspondence between the generated keywords and the visual subjects. Similar correspondences between the source phrases and the translated phrases are highlighted in attention-based neural machine translation models \cite{bahdanau2014neural}. While visual and textual data are easy to interpret and intuitive to humans, generic time-series data is more difficult to perceive, making the LSTM models a black box. This hinders the opportunity for post-training analysis. Few attempts have been made to employ attention functionality in LSTM in different time-series forecasting problems such as medial diagnosis \cite{choi2016retain} and weather forecast \cite{riemer2016correcting} or finance \cite{qin2017dual}. Although adding attention mechanism into recurrent architectures improves the performance and comprehensibility, it incurs high computational cost for the whole model. This impedes the practicality of the model in many financial forecasting situations in which the ability to rapidly train the system and make predictions with continuously large chunks of incoming data plays an important role.

Multivariate time-series data is naturally represented as a second-order tensor. This representation retains the temporal structure encoded in the data, which is essential for a learning model to capture temporal interaction and dependency. By applying a vectorization step, traditional vector-based models fail to capture such temporal cues, leading to inferior performance compared to tensor-based models that work directly on the natural representation of the input. Recent progress in mathematical tools and algorithms pertaining to tensor input have enabled the development of several learning systems based on multilinear projections. For example, popular discriminant and regression criteria were extended for tensor inputs, such as Multilinear Discriminant Analysis (MDA) \cite{li2014multilinear}, Multilinear Class-specific Discriminant Analysis (MCSDA) \cite{tran2017multilinear} or Tensor Regression (TR) \cite{guo2012tensor}. Regarding neural network formulations, attempts have also been made to learn separate projections for each mode of the input tensors \cite{gao2017matrix,do2017matrix}. Motivation to replace linear mapping by the multilinear counterpart stems from the fact that learning separate dependencies between separate modes of the input alleviates the so-called \textit{curse of dimensionality} and greatly reduces the amount of memory and computation required. While large volume of literature employing multilinear projections was developed for data modalities such as image, video, and text, few works have been dedicated to time-series prediction in general and financial time-series in particular.

In recent work \cite{tran2017tensor}, we have showed that a linear multivariate regression model could outperform other competing shallow architectures that do not take into account the temporal nature of HFT data. While performing reasonably well compared to other shallow architectures, the learning model in \cite{tran2017tensor} has certain short-comings in practice: the analytical solution is computed based on the entire dataset prohibiting its application in an online learning scenario; with large amount of data, this model clearly underfits the underlying generating process with performance inferior to other models based on deep architectures \cite{tsantekidis2017forecasting,tsantekidis2017using}. In this work, we propose a neural network layer which incorporates the idea of bilinear projection in order to learn two separate dependencies for the two modes of multivariate time-series data. Moreover, we augment the proposed layer with an attention mechanism that enables the layer to focus on important temporal instances of the input data. By formulation, the layer is differentiable. Thus it can be trained with any mini-batch gradient descend learning algorithm.

The contribution of our work is as follows:
\begin{itemize}
    \item We propose a new type of layer architecture for multivariate time-series data. The proposed layer is designed to leverage the idea of bilinear projection by incorporating an attention mechanism in the temporal mode. The formulation of the attention mechanism directly encourages the competition between neurons representing the same feature at different time instances. The learned model utilizing our proposed layer is highly interpretable by allowing us to look at which specific time instances the learning model attends to.
    \item We show both theoretically and experimentally that the proposed attention mechanism is highly efficient in terms of computational complexity, allowing development in practical financial forecasting systems.
    \item Numerical experiments in a large-scale Limit Order Book (LOB) dataset that contains more than 4 million limit orders show that by using a shallow architecture with only two hidden layers, we are able to outperform by a large margin existing results of deep networks, such as CNN and LSTM, leading to new state-of-the-art results. Furthermore, we show that our proposed attention mechanism can highlight the contribution of different temporal information, opening up opportunities for further analysis on the temporal instances of interest.
\end{itemize}

The rest of the paper is organized as follows: in section 2, we provide an overview of the related works focusing on time-series prediction problems. Section 3 describes our proposed layer architecture together with theoretical analysis on its computational complexity. Section 4 starts by describing the task of mid-price movement prediction given LOB data before providing details of our experimental procedures, results and quantitative analysis. Section 5 concludes our work and discusses possible future extension and analysis.

\section{Related Work}
Deep neural networks have been shown to be the state-of-the-art not only in human cognition tasks, such as language and image understanding but also in the prediction of complex time-series data. For example, RNN networks based on LSTM architectures have been used to predict the future rainfall intensity in different geographic areas \cite{xingjian2015convolutional}, commodity consumption \cite{riemer2016correcting} or to recognize patterns in clinical time-series \cite{lipton2015learning}. In financial data analysis, Deep Belief Networks and Auto-Encoders were used to derive portfolio trading models \cite{heaton2016deep,sharang2015using}. In addition, Deep Reinforcement Learning methods are also popular among the class of financial asset trading models \cite{deng2017deep,deng2015sparse}. Spatial relations between LOB levels was studied in \cite{sirignano2016deep} by a 3-hidden-layer multilayer perceptron (MLP) that models the joint distribution of bid and ask prices. Due to the erratic, noisy nature of stock price movement, many deep neural networks were proposed within a complex forecasting pipeline. For example, in high-frequency LOB data, the authors proposed to normalize the LOB states by the prior date's statistics before feeding them to a CNN \cite{tsantekidis2017forecasting} or an LSTM network \cite{tsantekidis2017using}. A more elaborate pipeline consisting of multi-resolution wavelet transform to filter the noisy input series, stacked Auto-Encoder to extract high-level representation of each stock index and an LSTM network to predict future prices was recently proposed in \cite{bao2017deep}. Along with popular deep networks, such as CNN, LSTM being applied to time-series forecasting problems, a recently proposed Neural Bag of Feature (NBoF) model was also applied to the problem of stock price movement prediction \cite{passalis2017time}. The architecture consists of an NBoF layer which compiles histogram representation of the input time-series and an MLP that classifies the extracted histograms. By learning the parameters of the whole network through back-propagation algorithm, NBoF was shown to outperform its Bag-of-Feature (BoF) counterpart.

In order to improve both performance and interpretability, attention mechanism was proposed for the recurrent sequence-to-sequence learning problem (ASeq-RNN) \cite{bahdanau2014neural}. Given a sequence of multivariate inputs $\mathbf{x}_i \in \mathbb{R}^{D}, 1 \leq i \leq T$ and an associated sequence of outputs $\mathbf{y}_j, 1 \leq j \leq T'$, the Seq-RNN model with attention mechanism learns to generate $\mathbf{y}_j$ from $\mathbf{x}_i$ by using three modules: the encoder, the memory and the decoder. The encoder maps each input $\mathbf{x}_i$ to a hidden state $\mathbf{h}_i^e$ using the nonlinear transformation $\mathbf{h}_{i}^{e} = f(\mathbf{x}_i, \mathbf{h}_{i-1}^e)$ coming from a recurrent unit (LSTM or GRU). From the sequence of hidden states generated by the encoder $\mathbf{h}_i^e, i=1,\dots,T$, the memory module generates context vectors $\mathbf{c}_j, j=1,\dots, T'$ for each output value $\mathbf{y}_j, j=1,\dots ,T'$. In a normal recurrent model, the context vector is simply selected as the last hidden state $\mathbf{h}_T^e$ while in the attention-based model, the context vectors provide summaries of the input sequence by linearly combining the hidden states $\mathbf{c}_{j} = \sum_{i=1}^{T} \alpha_{ij} \mathbf{h}_i^e$ through a set of attention weights $\alpha_{ij}$ learned by the following equations:

\begin{align}
e_{ij} &=\mathbf{v}_{\alpha}^{T} \tanh (\mathbf{W}_{\alpha}\mathbf{h}_{j-1}^d+\mathbf{U}_{\alpha}\mathbf{h}_i^e)\label{eq1} \\
\alpha_{ij} &=\frac{\exp (e_{ij})}{\sum_{k=1}^{T} \exp (e_{kj})}\label{eq2}
\end{align}

The softmax function in Eq. (\ref{eq2}) allows the model to produce the context vectors that focus on some time instances of the input sequence and $\mathbf{h}_{j}^d = g(\mathbf{y}_{j-1}, \mathbf{h}_{j-1}^d,\mathbf{c}_{j}), j=1,\dots , T'$ is the hidden state computed from the recurrent unit in the decoder module. From the hidden state $\mathbf{h}_j^d$ which is based on the previous state $\mathbf{h}_{j-1}^d$, the previous output $\mathbf{y}_{j-1}$ and the current context $\mathbf{c}_j$, the decoder learns to produce the output $\mathbf{y}_j = \mathbf{W}_{out} \mathbf{h}_j^d + \mathbf{b}_{out}$.

Based on the aforementioned attention mechanism, \cite{cinar2017position} proposed to replace Eq. (\ref{eq1}) with a modified attention calculation scheme that assumes the existence of pseudo-periods in the given time-series. Their experiments in energy consumption and weather forecast showed that the proposed model learned to attend particular time instances which indicate the pseudo-periods existing in the data. For the future stock price prediction task given current and past prices of multiple stock indices, the authors in \cite{qin2017dual} developed a recurrent network with two-stage attention mechanism which first focuses on different input series then different time instances. We should note that the formulations above of attention mechanism were proposed for the recurrent structure.

Our work can be seen as direct extension of \cite{tran2017tensor} in which the author proposed a regression model based on the bilinear mapping for the mid-price movement classification problem:
\begin{equation}\label{eq3}
f(\mathbf{X})=\mathbf{W}_1 \mathbf{X} \mathbf{w}_2
\end{equation}
where $\mathbf{X} \in \mathbb{R}^{D\times T}$ is a multivariate time-series containing $T$ temporal steps. $\mathbf{W}_1 \in \mathbb{R}^{3\times D}$ and $\mathbf{w}_2 \in \mathbb{R}^{T \times 1}$ are the parameters to estimate. By learning two separate mappings that transform the input LOB states to class-membership vector of size $3 \times 1$ corresponding to $3$ types of movements in mid-price, the regression model in \cite{tran2017tensor} was shown to outperform other shallow classifiers. Other related works that utilize a bilinear mapping function to construct a neural network layer include \cite{do2017matrix} and \cite{gao2017matrix}. While \cite{do2017matrix} attempted to incorporate the bilinear mapping into the recurrent structure by processing a block of temporal instances at each recurrent step, both \cite{do2017matrix} and \cite{gao2017matrix} focus on medium-scale visual-related tasks such as hand-written digit recognition, image interpolation and reconstruction.

\section{Proposed Method}

\subsection{Bilinear Layer}\label{BL}

We start this section by introducing some notations and definitions. Throughout the paper, we denote scalar values by either lower-case or upper-case character $(a,b,A,B,\dots)$, vectors by lower-case bold-face characters $(\mathbf{x}, \mathbf{y}, \dots)$, matrices by upper-case bold-face characters $(\mathbf{X}, \mathbf{Y},\dots )$. A matrix $\mathbf{X} \in \mathbb{R}^{D \times T}$ is a second order tensor which has two modes with $D$ and $T$ are the dimension of the first and second mode respectively. We denote $\mathbf{X}_i \in \mathbb{R}^{D \times T}, i=1, \dots, N$ the set of $N$ samples, each of which contains a sequence of $T$ past observations corresponding to its $T$ columns. The time span of the past values ($T$) is termed as \textit{history} while the time span in the future value ($H$) that we would like to predict is known as \textit{prediction horizon}. For example, given that the stock prices are sampled every second and $\mathbf{X}_i \in \mathbb{R}^{10 \times 100}$ contains stock prices at different LOB levels for the last $T=100$ seconds, the prediction horizon $H=10$ corresponds to predicting a future value, e.g. mid-price, at the next $10$ seconds.

Let us denote by $\mathbf{X} =[\mathbf{x}_{1},\dots , \mathbf{x}_{T_l}] \in \mathbb{R}^{D \times T}$ the input to the Bilinear Layer (BL). The layer transforms an input of size $D \times T$ to a matrix of size $D' \times T'$ by applying the following mapping:
\begin{equation}\label{eq4}
\mathbf{Y}= \phi \big( \mathbf{W}_{1} \mathbf{X} \mathbf{W}_{2} + \mathbf{B} \big)
\end{equation}
where $\mathbf{W}_{1} \in \mathbb{R}^{D'\times D}$ , $\mathbf{W}_{2}\in \mathbb{R}^{T \times T'}$, $\mathbf{B} \in \mathbb{R}^{D' \times T'}$ are the parameters to estimate. $\phi(\cdot)$ is an element-wise nonlinear transformation function, such as ReLU \cite{glorot2011deep} or sigmoid.

One of the obvious advantages of the mapping in Eq. (\ref{eq4}) is that the number of estimated parameters scales linearly with the dimension of each mode of the input rather than the number of input neurons. For an MLP layer, transforming an input of size $DT$ to $D'T'$ requires the estimation of $(DT+1)D'T'$ parameters (including the bias term), which are much higher than the number of parameters ($DD' + TT' + D'T'$) estimated by a bilinear layer.

A more important characteristic of the mapping in Eq. (\ref{eq4}), when it is applied to time-series data, is that the BL models two dependencies (one for each mode of the input representation), each of which has different semantic meanings. In order to better understand this, denote each column and row of $\mathbf{X}$ as $\mathbf{x}_{c_t} \in \mathbb{R}^{D}, t=1,\dots , T$ and $\mathbf{x}_{r_d} \in \mathbb{R}^{T}, d=1,\dots , D$, respectively. Given the input time-series $\mathbf{X}$, the $t$-th column represents $D$ different features or aspects of the underlying process observed at the time instance $t$, while the $d$-th row contains the temporal variations of the $d$-th feature during the past $T$ steps. Since
\begin{equation}\label{eq5}
\mathbf{W}_{1} \mathbf{X} = \big[ \mathbf{W}_{1} \mathbf{x}_{c_1}, \dots , \mathbf{W}_{1}\mathbf{x}_{c_{T}} \big]
\end{equation}
\begin{equation}\label{eq6}
\mathbf{X} \mathbf{W}_{2} = \begin{bmatrix}
(\mathbf{x}_{r_1})^{T} \mathbf{W}_{2} \\
\vdots \\
(\mathbf{x}_{r_{D}})^{T} \mathbf{W}_{2},
\end{bmatrix}
\end{equation}
Eq. (\ref{eq5}) shows that the interaction between different features/aspects at a time instance $t=1,\dots , T$ is captured by $\mathbf{W}_{1}$ while in Eq. (\ref{eq6}), $\mathbf{W}_{2}$ models the temporal progress of the $d$-th feature/aspect. For example, given that $\mathbf{X}$ contains stock prices of $D$ different LOB levels during the history $T$, the BL determines how different stock prices interact at a particular time instance by $\mathbf{W}_{1}$ and how the prices of a particular index progress over time by $\mathbf{W}_{2}$. It has been shown in \cite{sirignano2016deep} that taking advantage of the spatial structure existing in the LOB yields better joint distribution of the future best bid and ask prices.

\begin{figure}[t!]
	\includegraphics[width=\linewidth]{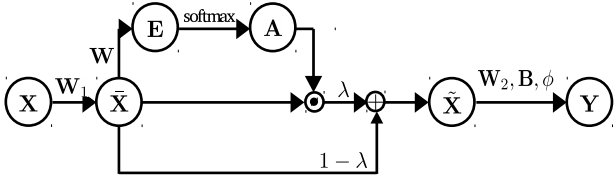}%
	\caption{Illustration of the proposed Temporal Attention augmented Bilinear Layer (TABL)}
	\label{TABL_layer}
\end{figure}

\subsection{Temporal Attention augmented Bilinear Layer}\label{TABL}
Although the BL learns separate dependencies along each mode, it is not obvious how a representation at a time instance interacts with other time instances or which time instances are important to the prediction at horizon $T'$. By incorporating the position information into the attention calculation scheme, the authors in \cite{cinar2017position} showed that the learned model only used a particular time instance in the past sequence to predict the future value at a given horizon for sequence-to-sequence learning. In order to learn the importance of each time instance in the proposed BL, we propose the Temporal Attention augmented Bilinear Layer (TABL) that maps the input $\mathbf{X} \in \mathbb{R}^{D \times T}$ to the output $\mathbf{Y} \in \mathbb{R}^{D' \times T'}$ as follows:
\begin{align}
\bar{\mathbf{X}} &= \mathbf{W}_{1} \mathbf{X}\label{eq7} \\
\mathbf{E} &= \bar{\mathbf{X}} \mathbf{W}\label{eq8} \\
\alpha_{ij} &= \frac{\exp (e_{ij})}{\sum_{k=1}^{T} \exp (e_{ik})}\label{eq9} \\
\tilde{\mathbf{X}} &= \lambda ( \bar{\mathbf{X}} \odot \mathbf{A}) + (1-\lambda)\bar{\mathbf{X}}\label{eq9_} \\
\mathbf{Y} &= \phi \big(  \tilde{\mathbf{X}} \mathbf{W}_{2} + \mathbf{B} \big)\label{eq10}
\end{align}
where $\alpha_{ij}$ and $e_{ij}$ denote the element at position $(i,j)$ of $\mathbf{A}$ and $\mathbf{E}$, respectively, $\odot$ denotes the element-wise multiplication operator and $\phi(\cdot)$ is a predefined nonlinear mapping as in Eq. (\ref{eq4}).

$\mathbf{W}_1 \in \mathbb{R}^{D' \times D}$, $\mathbf{W} \in \mathbb{R}^{T \times T}$, $\mathbf{W}_2 \in \mathbb{R}^{T \times T'}$, $\mathbf{B} \in \mathbb{R}^{D' \times T'}$ and $\lambda$ are the parameters of the proposed TABL. Similar to the aforementioned BL, TABL models two separate dependencies through $\mathbf{W}_1$ and $\mathbf{W}_2$ with the inclusion of the intermediate attention step learned through $\mathbf{W}$ and $\lambda$. The forward pass through TABL consists of $5$ steps, which are depicted in Figure \ref{TABL_layer}:

\begin{itemize}
\item In Eq. (\ref{eq7}), $\mathbf{W}_1$ is used to transform the representation of each time instance $\mathbf{x}_{c_t}, t=1,\dots , T$ of $\mathbf{X}$ (each column) to a new feature space $\mathbb{R}^{D'}$. This models the dependency along the first mode of $\mathbf{X}$ while keeping the temporal order intact.
\item The aim of the second step is to learn how important the temporal instances are to each other. This is realized by learning a structured matrix $\mathbf{W}$ whose diagonal elements are fixed to $1/T$. Let us denote by $\bar{\mathbf{x}}_t \in \mathbb{R}^{D'}$ and $\mathbf{e}_t \in \mathbb{R}^{D'}$ the $t$-th column of $\bar{\mathbf{X}}$ and $\mathbf{E}$ respectively. From Eq. (\ref{eq8}), we could see that $\mathbf{e}_t$ is the weighted combination of $T$ temporal instances in the feature space $\mathbb{R}^{D'}$, i.e. $T$ columns of $\bar{\mathbf{X}}$, with the weight of the $t$-th time instance always equal to $1/T$ since the diagonal elements of $\mathbf{W}$ are fixed to $1/T$. Thus, element $e_{ij}$ in $\mathbf{E}$ encodes the relative importance of element $\bar{x}_{ij}$ with respect to other $\bar{x}_{ik}, k \neq j$.
\item By normalizing the importance values in $\mathbf{E}$ using the softmax function in Eq. (\ref{eq9}), the proposed layer pushes many elements to become close to zero, while keeping the values of few of them positive. This process, produces the attention mask $\mathbf{A}$.
\item The attention mask $\mathbf{A}$ obtained from the third step is used to zero out the effect of unimportant elements in $\mathbb{R}^{D'}$. Instead of applying a \textit{hard} attention mechanism, the learnable scalar $\lambda$ in Eq. (\ref{eq9_}) allows the model to learn a \textit{soft} attention mechanism. In the early stage of the learning process, the learned features extracted from the previous layer can be noisy and might not be discriminative, thus \textit{hard} attention might mislead the model to unimportant information while \textit{soft} attention could enable the model to learn discriminative features in the early stage, i.e. before selecting the most important ones. Here we should note that $\lambda$ is constrained to lie in the range $[0,1]$, i.e. $0 \leq \lambda \leq 1$
\item Similar to BL, the final step of the proposed layer estimates the temporal mapping $\mathbf{W}_2$, extracting higher-level representation after the bias shift and nonlinearity transformation.
\end{itemize}

Generally, the introduction of attention mechanism in the second, third and fourth step of the proposed layer encourages the competition among neurons representing different temporal steps of the same feature, i.e. competition between elements on the same row of $\bar{\mathbf{X}}$. The competitions are, however, independent for each feature in $\mathbb{R}^{D'}$, i.e. elements of the same column of $\bar{\mathbf{X}}$ do not compete to be represented.

The proposed layer architecture is trained jointly with other layers in the network using the Back-Propagation (BP) algorithm. During the backward pass of BP, in order to update the parameters of TABL, the following quantities must be calculated: $\partial L/ \partial \mathbf{W}_1$, $\partial L/ \partial \mathbf{W}$, $\partial L/ \partial \lambda$, $\partial L/ \partial \mathbf{W}_2$ and $\partial L/ \partial \mathbf{B}$ with $L$ is the loss function. Derivation of these derivatives is given in the Appendix \ref{Ap1}.

\subsection{Complexity Analysis}
As mentioned in the previous section, the memory complexity of the BL is $O(DD'+TT'+D'T')$. The proposed TABL requires an additional amount of $O(T^2)$ in memory. Computation of the BL requires the following steps: matrix multiplication $\mathbf{W}_1 \mathbf{X} \mathbf{W}_2$ with the cost of $O(D'DT+D'TT')$, bias shift and nonlinear activation with the cost of $O(2D'T')$. In total, the computational complexity of BL is $O(D'DT+D'TT'+2D'T')$. Since TABL possesses the same computation steps as in BL with additional computation for the attention step, the total computational complexity of TABL is $O(D'DT+D'TT'+2D'T'+ D'T^{2}+3D'T)$ with the last two terms contributed from the applying the attention mask $\mathbf{A}$.

In order to compare our proposed temporal attention mechanism in bilinear structure with the attention mechanism in a recurrent structure, we estimate the complexity of the attention-based Seq-RNN (ASeq-RNN) proposed in \cite{bahdanau2014neural} as a reference. Let $D'$ denote the dimension of the hidden units in the encoder, memory and decoder module. In addition, we assume that input and output sequence have equal length. The total memory and computational complexity of ASeq-RNN are $O(3D'D+11D'^2+11D')$ and $O(11TD'^2+20TD'+4T^2D'+3TD'D+T^2)$ respectively. Details of the estimation are given in the Appendix \ref{Ap2}. While configurations of the recurrent and bilinear architecture are not directly comparable, it is still obvious that ASeq-RNN has much higher memory and computational complexity as compared to the proposed TABL. It should be noted that the given complexity of ASeq-RNN is derived based on GRU, which has lower memory and computation complexities compared to LSTM. Variants of ASeq-RNN proposed to time-series data are, however, based on LSTM units \cite{cinar2017position,qin2017dual}, making them even more computationally demanding.

\section{Experiments}
In this section, we evaluate our proposed architecture on the mid-price movement prediction problem based on a large-scale high-frequency LOB dataset. Before elaborating on experimental setups and numerical results, we start by describing the dataset and the prediction task.

\subsection{High-Frequency Limit Order Data}

In stock markets, traders buy and sell stocks through an order-driven system that aggregates all out-standing limit orders in limit order book. A limit order is a type of order to buy or sell a certain amount of a security at a specified price or better. In a limit order, the trader must specify type (buy/sell), price and the respective volume (number of stock items he/she wants to trade). Buy and sell limit orders constitute two sides of the Limit Order Book (LOB), i.e. the bid and ask side. At time $t$, the best bid price ($p_b^1(t)$) and best ask price ($p_a^1(t)$) are defined as the highest bid and lowest ask prices in the LOB, respectively. When a new limit order arrives, the LOB aggregates and sorts the orders on both sides based on the given prices so that best bid and best ask price are placed at the first level. If there are limit orders where the bid price is equal or higher than the lowest ask, i.e. $p_b^1(t) \geq p_a^1(t)$, those orders are immediately fulfilled and removed from the orders book. In contrast to limit orders, a buy market order is executed immediately with the current best ask price while a sell market order is executed at the current best bid price. An arriving market order is immediately matched with the best available price in the limit order book and a trade occurs, which decreases the depth of the LOB by the amount of shares traded. There are multiple price levels in the order book and in this paper we consider 10 top price levels from both sides of the LOB. For more information on limit order books, we refer \cite{cont2011statistical}.

The LOB reflects the existing supply and demand for the stock at different price levels. Therefore, based on the availability of LOB data, several analysis and prediction problems can be formulated such as modeling the order flow distribution, the joint distribution of best bid and ask price or casualty analysis of turbulence in price change. The mid-price at a given time instance is a quantity defined as the mean between the best bid price and best ask price:
\begin{equation}\label{eq11}
p_t = \frac{p_a^1(t) + p_b^1(t)}{2}
\end{equation}
This quantity is a virtual price since no trade can take place at this exact price. Since this quantity lies in the middle of the best bid and best ask price, its movement reflects the dynamics of LOB and the market. Therefore, being able to predict the mid-price changes in the future is of great importance.

We evaluate our proposed architecture with the task of predicting the future movement of mid-price given the past bid and ask prices with respective volumes. We use the publicly available dataset provided in \cite{ntakaris2017benchmark}, known as FI-2010 dataset. The data were collected from 5 different Finnish stocks in NASDAQ Nordic coming from different industrial sectors. The collection period is from 1st of June to 14th of June 2010, producing order data of 10 working days with approximately 4.5 million events. For each event, the prices and volumes of the top 10 orders from each side of the LOB were extracted, producing a $40$-dimensional vector representation. In \cite{ntakaris2017benchmark}, the author extracts a $144$-dimensional feature vector for every non-overlapping block of 10 events, with the first $40$ dimensions containing the prices and volumes of the last event in the block, while the rest contain extracted information within the block. The feature extraction process results in $453,975$ feature vectors in total. For each feature vector, the dataset includes labels of the mid-price movement (stationary, increase, decrease) in $5$ different horizons ($H={10,20,30,50,100}$) corresponding to the future movements in the next $10,20,30,50,100$ events.

\begin{table}[t!]
	\begin{center}
		\caption{Experiment Results in Setup1}\label{t1}
		\resizebox{\linewidth}{!}{
			\begin{tabular}{|c|c|c|c|c|}
				\multicolumn{5}{c}{} \\ \hline
				\textbf{Models}		& \textbf{Accuracy \%} 	& \textbf{Precision \%} & \textbf{Recall \%}	& \textbf{F1 \%} 		\\ \hline \hline
				\multicolumn{5}{|c|}{\textit{Prediction Horizon $H=10$}} \\ \hline
				RR\cite{ntakaris2017benchmark}		& $48.00$	& $41.80$	& $43.50$	& $41.00$	\\ \hline		
				SLFN\cite{ntakaris2017benchmark}	& $64.30$	& $51.20$	& $36.60$	& $32.70$	\\ \hline
				LDA\cite{tran2017tensor}		& $63.82$	& $37.93$	&$45.80$	& $36.28$	\\ \hline
				MDA\cite{tran2017tensor}		& $71.92$	& $44.21$	&$60.07$	& $46.06$	\\ \hline
				MCSDA\cite{tran2017multilinear}	& $83.66$	& $46.11$	&$48.00$	& $46.72$ \\ \hline
				MTR\cite{tran2017tensor}		& $86.08$	& $51.68$	&$40.81$	& $40.14$	\\ \hline	
				WMTR\cite{tran2017tensor}	& $81.89$	& $46.25$	&$51.29$	& $47.87$	\\ \hline
				BoF\cite{passalis2017time}		& $57.59$	& $39.26$	&$51.44$	& $36.28$	\\ \hline
				N-BoF\cite{passalis2017time}	& $62.70$	& $42.28$	&$61.41$	& $41.63$	\\ \hline \hline
				
				A(BL)	& $44.48$	& $47.56$	&$50.78$	& $43.05$ \\ \hline
				\textbf{A(TABL)}	& $66.03$	& $56.48$	&$58.09$	& $\mathbf{56.50}$ \\ \hline \hline
				
				B(BL)	& $72.80$	& $65.25$	&$66.92$	& $65.59$ \\ \hline
				\textbf{B(TABL)}	& $73.62$	& $66.16$	&$68.81$	& $\mathbf{67.12}$ \\ \hline \hline
				
				C(BL)	& $76.82$	& $70.51$	&$72.75$	& $71.33$ \\ \hline
				\textbf{C(TABL)}	& $78.01$	& $72.03$	&$74.06$	& $\mathbf{72.84}$ \\ \hline \hline
				
				\multicolumn{5}{|c|}{\textit{Prediction Horizon $H=50$}} \\ \hline
				RR\cite{ntakaris2017benchmark}		& $43.90$	& $43.60$	& $43.30$	& $42.70$	\\ \hline		
				SLFN\cite{ntakaris2017benchmark}	& $47.30$	& $46.80$	& $46.40$	& $45.90$	\\ \hline
				BoF\cite{passalis2017time}		& $50.21$	& $42.56$	&$49.57$	& $39.56$	\\ \hline
				N-BoF\cite{passalis2017time}	& $56.52$	& $47.20$	&$58.17$	& $46.15$	\\ \hline \hline
				
				A(BL)	& $46.47$	& $54.58$	&$47.83$	& $44.51$ \\ \hline
				\textbf{A(TABL)}	& $54.61$	& $54.89$	&$53.13$	& $\mathbf{53.00}$ \\ \hline \hline
				
				B(BL)	& $68.09$	& $67.95$	&$67.12$	& $67.16$ \\ \hline
				\textbf{B(TABL)}	& $69.54$	& $69.12$	&$68.84$	& $\mathbf{68.84}$ \\ \hline \hline
				
				C(BL)	& $74.46$	& $74.20$	&$73.95$	& $73.79$ \\ \hline
				\textbf{C(TABL)}	& $74.81$	& $74.58$	&$74.27$	& $\mathbf{74.32}$ \\ \hline
				\hline
				
				\multicolumn{5}{|c|}{\textit{Prediction Horizon $H=100$}} \\ \hline
				RR\cite{ntakaris2017benchmark}		& $42.90$	& $42.90$	& $42.90$	& $41.60$	\\ \hline		
				SLFN\cite{ntakaris2017benchmark}	& $47.70$	& $45.30$	& $43.20$	& $41.00$	\\ \hline				
				BoF\cite{passalis2017time}		& $50.97$	& $42.48$	&$47.84$	& $40.84$	\\ \hline
				N-BoF\cite{passalis2017time}	& $56.43$	& $47.27$	&$54.99$	& $46.86$	\\ \hline \hline
				
				A(BL)	& $48.90$	& $53.23$	&$45.41$	& $43.40$ \\ \hline
				\textbf{A(TABL)}	& $51.35$	& $51.37$	&$52.02$	& $\mathbf{50.66}$ \\ \hline \hline
				
				B(BL)	& $66.02$	& $65.78$	&$66.63$	& $65.60$ \\ \hline
				\textbf{B(TABL)}	& $69.31$	& $68.95$	&$69.41$	& $\mathbf{68.86}$ \\ \hline \hline
				
				C(BL)	& $73.80$	& $73.43$	&$73.40$	& $73.21$ \\ \hline
				\textbf{C(TABL)}	& $74.07$	& $73.51$	&$73.80$	& $\mathbf{73.52}$ \\ \hline
			\end{tabular}
		}
	\end{center}
\end{table}

There exist two experimental setups using FI-2010 dataset. The first setting is the standard anchored forward splits provided by the database which we will refer as Setup1. In Setup1, the dataset is divided into $9$ folds based on a day basis. Specifically, in the $k$-th fold, data from the first $k$ days is used as the train set while data in the ($k+1$)-th day is used as a test set with $k=1,\dots , 9$. The second setting, referred as Setup2, comes from recent works \cite{tsantekidis2017forecasting,tsantekidis2017using} in which deep network architectures were evaluated. In Setup2, the first $7$ days are used as the train set while the last $3$ days are used as test set. We provide the evaluation of our proposed architecture in both settings using the z-score normalized data provided by the database.

\subsection{Network Architecture}\label{architecture}

\begin{figure}[t!]
	\centering
	\includegraphics[width=0.9\linewidth]{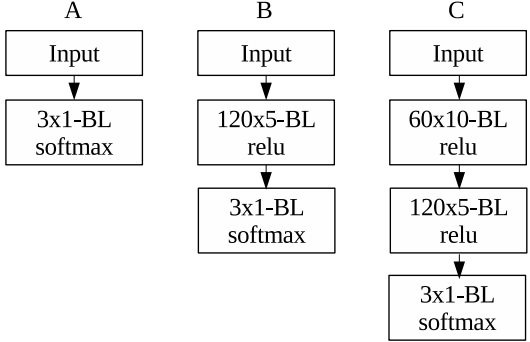}
	\caption{Baseline Network Topologies}
	\label{topology}
\end{figure}

In order to evaluate the bilinear structure in general and the proposed Temporal Attention augmented Bilinear Layer (TABL) in particular, we construct three different baseline network configurations (A,B,C) with $d=\{0,1,2\}$ hidden layers that are all Bilinear Layer (BL). Details of the baseline network configurations are shown in Figure \ref{topology}. The input to all configurations is a matrix of size $40\times 10$ which contains prices and volumes of the top $10$ orders from bid and ask side ($40$ values) spanning over a history of $100$ events\footnote{Since the feature vector is extracted from a block of 10 events and we only use the first 40 dimensions of the given feature vector, which correspond to prices and volumes of the last event in the block}. Here $120\times5$-BL denotes the Bilinear Layer with output size $120\times 5$. Based on the baseline network configurations, hereby referred as A(BL), B(BL) and C(BL), we replace the last BL classification layer by the proposed attention layer (TABL) to evaluate the effectiveness of attention mechanism. The resulting attention-based configurations are denoted as A(TABL), B(TABL) and C(TABL). Although attention mechanism can be placed in any layer, we argue that it is more beneficial for the network to attend to high-level representation, which is similar to visual attention mechanism that is applied after applying several convolution layers \cite{xu2015show}. In our experiments, we made no attempt to validate all possible positions to apply attention mechanism by simply incorporating it into the last layer.

\begin{table}[t!]
	\begin{center}
		\caption{Experiment Results in Setup2}\label{t2}
		\resizebox{\linewidth}{!}{
			\begin{tabular}{|c|c|c|c|c|}
				\multicolumn{5}{c}{} \\ \hline
				\textbf{Models}		& \textbf{Accuracy \%} 	& \textbf{Precision \%} & \textbf{Recall \%}	& \textbf{F1 \%} 		\\ \hline \hline
				\multicolumn{5}{|c|}{\textit{Prediction Horizon $H=10$}} \\ \hline
				SVM\cite{tsantekidis2017using}		& -			& $39.62$	&$44.92$	& $35.88$	\\ \hline	
				MLP\cite{tsantekidis2017using}		& -			& $47.81$	&$60.78$	& $48.27$	\\ \hline
				CNN\cite{tsantekidis2017forecasting}		& -			& $50.98$	&$65.54$	& $55.21$	\\ \hline
				LSTM\cite{tsantekidis2017using}	& -			& $60.77$	&$75.92$	& $66.33$	\\ \hline \hline
				
				A(BL)	& $29.21$	& $44.08$	&$48.14$	& $29.47$ \\ \hline
				\textbf{A(TABL)}	& $70.13$	& $56.28$	&$58.26$	& $\mathbf{56.03}$ \\ \hline \hline
				
				B(BL)	& $78.37$	& $67.73$	&$68.89$	& $67.71$ \\ \hline
				\textbf{B(TABL)}	& $78.91$	& $68.04$	&$71.21$	& $\mathbf{69.20}$ \\ \hline \hline
				
				C(BL)	& $82.52$	& $73.89$	&$76.22$	& $75.01$ \\ \hline
				\textbf{C(TABL)}	& $84.70$	& $76.95$	&$78.44$	& $\mathbf{77.63}$ \\ \hline \hline
				
				\multicolumn{5}{|c|}{\textit{Prediction Horizon $H=20$}} \\ \hline
				SVM\cite{tsantekidis2017using}		& -			& $45.08$	&$47.77$	& $43.20$	\\ \hline	
				MLP\cite{tsantekidis2017using}		& -			& $51.33$	&$65.20$	& $51.12$	\\ \hline
				CNN\cite{tsantekidis2017forecasting}		& -			& $54.79$	&$67.38$	& $59.17$	\\ \hline
				LSTM\cite{tsantekidis2017using}	& -			& $59.60$	&$70.52$	& $62.37$	\\ \hline \hline
				
				A(BL)	& $42.01$	& $47.71$	&$45.38$	& $38.61$ \\ \hline
				\textbf{A(TABL)}	& $62.54$	& $52.36$	&$50.96$	& $\mathbf{50.69}$ \\ \hline \hline
				
				B(BL)	& $70.33$	& $62.97$	&$60.64$	& $61.02$ \\ \hline
				\textbf{B(TABL)}	& $70.80$	& $63.14$	&$62.25$	& $\mathbf{62.22}$ \\ \hline \hline
				
				C(BL)	& $72.05$	& $65.04$	&$65.23$	& $64.89$ \\ \hline
				\textbf{C(TABL)}	& $73.74$	& $67.18$	&$66.94$	& $\mathbf{66.93}$ \\ \hline \hline
				
				\multicolumn{5}{|c|}{\textit{Prediction Horizon $H=50$}} \\ \hline
				SVM\cite{tsantekidis2017using}		& -			& $46.05$	&$60.30$	& $49.42$	\\ \hline	
				MLP\cite{tsantekidis2017using}		& -			& $55.21$	&$67.14$	& $55.95$	\\ \hline
				CNN\cite{tsantekidis2017forecasting}		& -			& $55.58$	&$67.12$	& $59.44$	\\ \hline
				LSTM\cite{tsantekidis2017using}	& -			& $60.03$	&$68.58$	& $61.43$	\\ \hline \hline
				
				A(BL)	& $51.92$	& $51.59$	&$50.35$	& $49.58$ \\ \hline
				\textbf{A(TABL)}	& $60.15$	& $59.05$	&$55.71$	& $\mathbf{55.87}$ \\ \hline \hline
				
				B(BL)	& $72.16$	& $71.28$	&$68.69$	& $69.40$ \\ \hline
				\textbf{B(TABL)}	& $75.58$	& $74.58$	&$73.09$	& $\mathbf{73.64}$ \\ \hline \hline
				
				C(BL)	& $78.96$	& $77.85$	&$77.04$	& $77.40$ \\ \hline
				\textbf{C(TABL)}	& $79.87$	& $79.05$	&$77.04$	& $\mathbf{78.44}$ \\ \hline
			\end{tabular}
		}
	\end{center}
\end{table}

\subsection{Experiment Settings}
The following experimental settings were applied to all network configurations mentioned in the previous subsection. We have experimented by training the networks with two types of stochastic optimizers: SGD \cite{sutskever2013importance} and Adam \cite{kingma2014adam}. For SGD, the Nesterov momentum was set to $0.9$ while for Adam, the exponential decay rates of the first and second moment were fixed to $0.9$ and $0.999$ respectively. The initial learning rate of both optimizers was set to $0.01$ and decreased by the following learning rate schedule $SC=\{0.01,0.005, 0.001, 0.0005, 0.0001\}$ when the loss in the training set stops decreasing. In total, all configurations were trained for maximum $200$ epochs with the mini-batch size of $256$ samples.

Regarding regularization techniques, we used a combination of dropout and max-norm \cite{srivastava2014dropout}, which was shown to improve generalization capacity of the network. Dropout was applied to the output of all hidden layers with a fixed percentage of  $0.1$. Max-norm regularizer is a type of weight constraint that enforces an absolute upper bound on the $l_2$ norm of the incoming weights to a neuron. The maximum norm was validated from the set $\{3.0, 5.0, 7.0\}$. Although weight decay is a popular regularization technique in deep neural network training, our exploratory experiments indicated that weight decay is not a suitable regularization option when training the bilinear structure.

We followed similar approach proposed in \cite{tran2017tensor} to weight the contribution of each class in the loss function. Since the evaluated network structures output the class-membership probability vector, the weighted entropy loss function was used:

\begin{figure*}[t!]
	\makebox[\textwidth][c]{\includegraphics[width=1.2\textwidth]{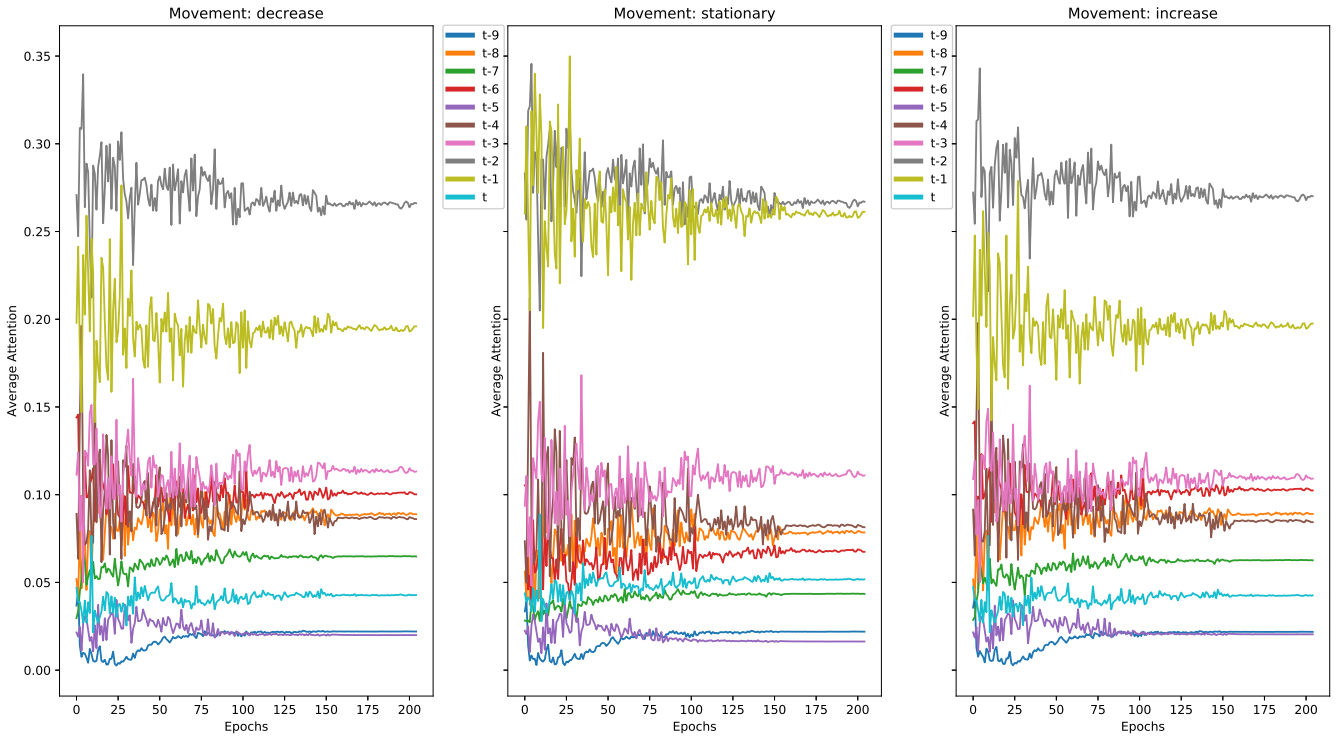}}%
	\caption{Average attention of 10 temporal instances during training in 3 types of movement: decrease, stationary, increase. Values taken from configuration A(TABL) in Setup2, horizon $H=10$}
	\label{attention}
\end{figure*}

\begin{equation}\label{eq12}
L=- \sum_{i=1}^{3} \frac{c}{N_i}y_i \log(\tilde{y}_i)
\end{equation}
where $N_i$, $y_i$, $\tilde{y}_i$ are the number of samples, true probability and the predicted probability of the $i$-th class respectively. $c=1e6$ is a constant used to ensure numerical stability by avoiding the loss values being too small when dividing by $N_i$.

All evaluated networks were initialized with the random initialization scheme proposed in \cite{he2015delving} except the attention weight $\mathbf{W}$ and $\lambda$ of the TABL layer. Randomly initializing $\mathbf{W}$ might cause the layer to falsely attend to unimportant input parts, leading the network to a bad local minima. We thus initialized $\lambda=0.5$ and all elements in $\mathbf{W}$ by a constant equal to $1/T$ with $T$ is input dimension of the second mode. By initializing $\mathbf{W}$ with a constant, we ensure that the layer starts by putting equal focus on all temporal instances.

\subsection{Experiment Results}
Following the experimental settings detailed in the previous subsection, we evaluated the proposed network structures in both Setup1 and Setup2. Besides the performance of our proposed network structures, we also report here all available experiment results coming from different models including Ridge Regression (RR), Single-Layer-Feedforward Network (SLFN), Linear Discriminant Analysis (LDA), Multilinear Discriminant Analysis (MDA), Multilinear Time-series Regression (MTR), Weighted Multilinear Time-series Regression (WMTR) \cite{tran2017tensor}, Multilinear Class-specific Discriminant Analysis (MCSDA) \cite{tran2017multilinear}, Bag-of-Feature (BoF), Neural Bag-of-Feature (N-BoF) \cite{passalis2017time} in Setup1 and Support Vector Machine (SVM), Multilayer Perceptron (MLP), Convolutional Neural Network (CNN) \cite{tsantekidis2017forecasting} and LSTM \cite{tsantekidis2017using} in Setup2.


Since the dataset is unbalanced with the majority of samples belonging to the stationary class, we tuned the hyper-parameters based on the average F1 score per class, which is a trade-off between precision and recall, measured on the training set. With the optimal parameter setting, the average performance on the test set over $9$ folds is reported in Setup1 while in Setup2, each network configuration is trained $5$ times and the average performance on the test set over $5$ runs is reported. Besides the main performance metric F1, we also report the corresponding accuracy, average precision per class and average recall, also known as sensitivity, per class.

Table \ref{t1} and \ref{t2} report the experiment results in Setup1 and Setup2, respectively. As can be seen in Table \ref{t1}, all the competing models in Setup1 belong to the class of shallow architectures with maximum 2 hidden layers (C(BL), C(TABL) and N-BoF). It is clear that all of the bilinear structures outperform other competing models by a large margin for all prediction horizons with the best performances coming from bilinear networks augmented with attention mechanism. Notably, average F1 obtained from the 2 hidden-layer configuration with TABL exceeds the previous best result in Setup1 achieved by WMTR in \cite{tran2017tensor} by nearly $25$\%. Although NBoF and C(TABL) are both neural network-based architectures with 2 hidden layers, C(TABL) surpasses NBoF by nearly $30$\% on all horizons. This is not surprising since a regression model based on bilinear projection was shown to even outperform NBoF in \cite{tran2017tensor}, indicating that by separately learning dependencies in different modes is crucial in time-series LOB data prediction.

While experiments in Setup1 show that the bilinear structure in general and the proposed attention mechanism in particular outperform all of the existing models that exploit shallow architectures, experiments in Setup2 establish the comparison between conventional deep neural network architectures and the proposed shallow bilinear architectures. Even with 1 hidden-layer, TABL performs similarly ($H=20$) or largely better than the previous state-of-the-art results obtained from LSTM network ($H=10,50$). Although being deep with 7 hidden layers, the CNN model is greatly inferior to the proposed ones. Here we should note that the CNN proposed in \cite{tsantekidis2017forecasting} gradually extracts local temporal information by the convolution layers. On the other hand, the evaluated bilinear structures fuse global temporal information from the beginning, i.e. the first layer. The comparison between CNN model and bilinear ones might indicate that the global temporal cues learned in the later stage of CNN (after some convolution layers) lose the discriminative global information existing in the raw data.

\begin{table}[t!]
	\begin{center}
		\caption{Average Computation Time Of State-Of-The-Art Models}\label{t3}
		\resizebox{\linewidth}{!}{
			\begin{tabular}{|c|c|c|c|}
				\multicolumn{4}{c}{} \\ \hline
				\textbf{Models} & \textbf{Forward} (ms)	& \textbf{Backward} (ms) & \textbf{Total} (ms)	\\ \hline
				C(BL)		& $0.0253$	&$0.0327$	& $0.0580$	\\ \hline	
				C(TABL)		& $0.0254$	&$0.0344$	& $0.0598$	\\ \hline
				CNN		& $0.0613$	&$0.1100$	& $0.1713$	\\ \hline
				LSTM	& $0.2291$	&$0.3487$	& $0.5778$	\\ \hline
			\end{tabular}
		}
	\end{center}
\end{table}

\begin{figure}[t!]
	\centering
	\includegraphics[width=\linewidth]{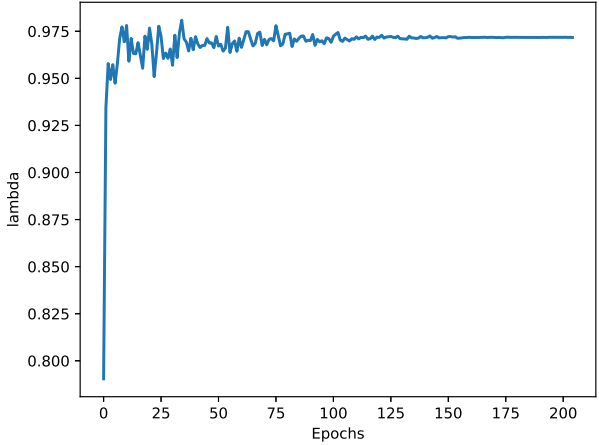}
	\caption{Corresponding $\lambda$ during training in A(TABL) in Setup2 and horizon $H=10$}\label{lambda}
\end{figure}

Comparing BL and TABL, it is clear that adding the attention mechanism improves the performance of the bilinear networks with only small increase in the number of parameters. More importantly, the attention mechanism opens up opportunities for further analyzing the contribution of the temporal instances being attended to. This can be done by looking into the attention mask $\mathbf{A}$. During the training process, each element in $\mathbf{A}$ represents the amount of attention the corresponding element in $\bar{\mathbf{X}}$ receives. In order to observe how each of the $10$ events in the input data contributes to the decision function, we analyze the statistics during the training process of the configuration A(TABL) in Setup2 with horizon $H=10$. Figure \ref{attention} plots the average attention values of each column of $\mathbf{A}$ which correspond to the average attention the model gives to each temporal instance during the training process. The three plots correspond to attention patterns in three types of movement of the mid-price. It is clear that the given model focuses more on some events such as the second ($t-1$), third ($t-2$) and fourth ($t-3$) most recent event in all types of movement. While the attention patterns are similar for the decrease and increase class, they are both different when compared to those of the stationary class. This indicates that when the mid-price is going to move from its equilibrium, the model can shift its attention to different events in order to detect the prospective change. Figure \ref{lambda} shows the corresponding values of $\lambda$ from the same model after every epoch during the training process. As can be seen from Figure \ref{lambda}, $\lambda$ increases in the first few steps before stabilizing close to $1$, which illustrates a \textit{soft} attention behavior achieved by $\lambda$ described in section \ref{TABL}. The insights into the attention patterns and the amount of attention received by each event given by the proposed attention-based layer could facilitate further quantitative analysis such as casualty or pseudo-period analysis.

Table \ref{t3} reports the average computation time of C(BL), C(TABL), CNN \cite{tsantekidis2017forecasting}, LSTM \cite{tsantekidis2017using} measured on the same machine with CPU core i7-4790 and 32 GB of memory. The second, third and last column shows the average time (in millisecond) taken by the forward pass, backward pass and one training pass of a single sample in the state-of-the-art models. It is obvious that the proposed attention mechanism only increases the computational cost by a relatively small margin. On the contrary, previously proposed deep neural network architectures require around $3\times$ (CNN) and $10\times$ longer to train or make inference while having inferior performances compared to the proposed architecture. This points out that our proposed architecture excels previous best models not only in performance but also in efficiency and practicality in applications such as High-Frequency Trading.

\section{Conclusions}
In this paper, we proposed a new neural network layer type for multivariate time-series data analysis. The proposed layer leverages the idea of bilinear projection and is augmented with a temporal attention mechanism. We provide theoretical analysis on the complexity of the proposed layer in comparison with existing attention mechanisms in recurrent structure, indicating that the proposed layer possesses much lower memory and computational complexity. Extensive experiments in a large-scale Limit Order Book dataset show the effectiveness of the proposed architecture: with only $2$ hidden-layers, we can surpass existing state-of-the-art models by a large margin. The proposed temporal attention mechanism not only improves the performance of the bilinear structure but also enhances its interpretability. Our quantitative analysis of the attention patterns during the training process opens up opportunities in future research of the patterns of interest.

\appendices
\section{TABL Derivatives}\label{Ap1}
In order to calculate the derivatives of TABL, we follow the notation: given $\mathbf{X} \in \mathbb{R}^{I\times J}$ and $\mathbf{X} \in \mathbb{R}^{M \times N}$, $\partial \mathbf{Y}/ \partial \mathbf{X}$ is a matrix of size $IJ \times MN$ with element at $(ij,mn)$ equal to $\partial \mathbf{Y}_{ij}/ \partial \mathbf{B}_{mn}$. Similarly $\partial L /\partial \mathbf{X} \in \mathbb{R}^{1 \times MN}$ with $L \in \mathbb{R}, \mathbf{X} \in \mathbb{R}^{M \times N}$. Denote $\mathbf{I}_M \in \mathbb{R}^{M \times M}$ the identity matrix and $\mathbf{1}_{MN} \in \mathbb{R}^{M \times N}$ a matrix with all elements equal to $1$. In addition, our derivation heavily uses the following formulas:
\begin{align}\label{eqB1}
\frac{\partial (\mathbf{A}\mathbf{X}\mathbf{B})}{\partial \mathbf{X}} & = \mathbf{B}^{T} \otimes \mathbf{A} \\
\frac{\partial (\mathbf{A} \odot \mathbf{B})}{\partial \mathbf{C}} &= diag \big( vec (\mathbf{A}) \big)  \odot \frac{\partial \mathbf{B}}{\partial \mathbf{C}} \nonumber \\
& + diag \big( vec (\mathbf{B}) \big)  \odot \frac{\partial \mathbf{A}}{\partial \mathbf{C}}
\end{align}
where $\otimes$ denotes the Kronecker product, $vec(\mathbf{A})$ denotes the vectorization operator that concatenates columns of $\mathbf{A}$ into one vector and $diag(\mathbf{x})$ denotes the diagonal matrix with the diagonal elements taken from $\mathbf{x}$.

We proceed by calculating the derivate of the left-hand side with respect to every term on the right-hand side from Eq. (\ref{eq7}) to (\ref{eq10}):
\begin{itemize}
\item From Eq. (\ref{eq7})
\begin{align}
\frac{\partial \bar{\mathbf{X}}}{\partial \mathbf{W}_1} &= \frac{\partial (\mathbf{I}_{D'} \mathbf{W}_1 \mathbf{X})}{\partial \mathbf{W}_1} = \mathbf{X}^{T} \otimes \mathbf{I}_{D'} \\
\frac{\partial \bar{\mathbf{X}}}{\partial \mathbf{X}} &= \frac{\partial (\mathbf{W}_1 \mathbf{X} \mathbf{I}_{T})}{\partial \mathbf{W}_1} = \mathbf{I}_{T} \otimes \mathbf{W}_1
\end{align}

\item From Eq. (\ref{eq8})
\begin{align}
\frac{\partial \mathbf{E}}{\partial \bar{\mathbf{X}}} &= \frac{\partial (\mathbf{I}_{D'} \bar{\mathbf{X}} \mathbf{W})}{\partial \bar{\mathbf{X}}} =  \mathbf{W}^{T} \otimes \mathbf{I}_{D'} \\
\frac{\partial \mathbf{E}}{\partial \mathbf{W}} &= \frac{\partial (\bar{\mathbf{X}} \mathbf{W} \mathbf{I}_{T})}{\partial \bar{\mathbf{X}}} =  \mathbf{I}_{T} \otimes \bar{\mathbf{X}}
\end{align}

\item From Eq. (\ref{eq9}) $\partial \mathbf{A}/{\partial \mathbf{E}}$ is calculated by the following results:
\begin{align}
\frac{\partial \alpha_{ij}}{\partial e_{ij}} &= \alpha_{ij}- \alpha_{ij}^2  , \forall i,j \\
\frac{\partial \alpha_{ij}}{\partial e_{ip}} &= \frac{\alpha_{ij}(1- \alpha_{ij})}{\sum_{k \neq j,p} \exp (e_{ik})} , \forall p \neq j \\
\frac{\partial \alpha_{ij}}{\partial e_{pq}} &= 0 , \forall p \neq i
\end{align}

\item From Eq. (\ref{eq9_})
\begin{align}
\frac{\partial \tilde{\mathbf{X}}}{\partial \mathbf{A}} &= \lambda \frac{\partial (\bar{\mathbf{X}} \odot \mathbf{A})}{\partial \mathbf{A}} =  \lambda diag\big( vec(\bar{\mathbf{X}})\big) \\
\frac{\partial \tilde{\mathbf{X}}}{\partial \bar{\mathbf{X}}} &= \frac{\partial ( [\lambda \mathbf{A}+(1-\lambda )\mathbf{1}_{D'T}] \odot \bar{\mathbf{X}})}{\partial \bar{\mathbf{X}}} \nonumber\\
 &=   diag\big( vec(\lambda \mathbf{A}+(1-\lambda )\mathbf{1}_{D'T})\big)\nonumber \\
 &+ diag \big( vec(\bar{\mathbf{X}})\big) \odot \big( \lambda \frac{\partial \mathbf{A}}{\partial \bar{\mathbf{X}}}\big) \nonumber\\
 &=  diag\big( vec(\lambda \mathbf{A}+(1-\lambda )\mathbf{1}_{D'T})\big) \nonumber\\
 &+ diag \big( vec(\bar{\mathbf{X}})\big) \odot \big( \frac{\partial \mathbf{A}}{\partial \mathbf{E}} \frac{\partial \mathbf{E}}{\partial \bar{\mathbf{X}}}\big)\\
\frac{\partial \tilde{\mathbf{X}}}{\partial \lambda} &= (\bar{\mathbf{X}} \odot \mathbf{A}-\bar{\mathbf{X}})
\end{align}

\item In Eq. (\ref{eq10}), denote $\bar{\mathbf{Y}}= \tilde{\mathbf{X}}\mathbf{W}_2 + \mathbf{B}$, Eq. (\ref{eq10}) becomes $\mathbf{Y} = \phi (\bar{\mathbf{Y}})$ and we have:
\begin{align}
\frac{\partial \mathbf{Y}}{\partial \tilde{\mathbf{X}}} &= \frac{\partial \mathbf{Y}}{\partial \bar{\mathbf{Y}}} \frac{\partial \bar{\mathbf{Y}}}{\partial \tilde{\mathbf{X}}} = \frac{\partial \phi (\bar{\mathbf{Y}})}{\partial \bar{\mathbf{Y}}} \frac{\partial (\mathbf{I}_{D'}\tilde{\mathbf{X}}\mathbf{W}_2)}{\partial \tilde{\mathbf{X}}} \nonumber\\
&=\frac{\partial \phi (\bar{\mathbf{Y}})}{\partial \bar{\mathbf{Y}}}  \mathbf{W}_{2}^{T} \otimes \mathbf{I}_{D'}  \\
\frac{\partial \mathbf{Y}}{\partial \mathbf{W}_2} &=\frac{\partial \mathbf{Y}}{\partial \bar{\mathbf{Y}}} \frac{\partial \bar{\mathbf{Y}}}{\partial \mathbf{W}_2} = \frac{\partial \phi (\bar{\mathbf{Y}})}{\partial \bar{\mathbf{Y}}}\frac{\partial (\tilde{\mathbf{X}}\mathbf{W}_2\mathbf{I}_{T})}{\partial \mathbf{W}_2} \nonumber\\
&= \frac{\partial \phi (\bar{\mathbf{Y}})}{\partial \bar{\mathbf{Y}}}\mathbf{I}_{T} \otimes \tilde{\mathbf{X}} \\
\frac{\partial \mathbf{Y}}{\partial \mathbf{B}} &=\frac{\partial \mathbf{Y}}{\partial \bar{\mathbf{Y}}} \frac{\partial \bar{\mathbf{Y}}}{\partial \mathbf{B}} = \frac{\partial \phi (\bar{\mathbf{Y}})}{\partial \bar{\mathbf{Y}}}\frac{\partial (\mathbf{I}_{D'}\mathbf{B}\mathbf{I}_{T})}{\partial \mathbf{B}} \nonumber\\
&= \frac{\partial \phi (\bar{\mathbf{Y}})}{\partial \bar{\mathbf{Y}}}\mathbf{I}_{T} \otimes \mathbf{I}_{D'}
\end{align}
where $\partial \phi (\bar{\mathbf{Y}})/{\partial \bar{\mathbf{Y}}}$ is the derivative of the element-wise activation function, which depends on the form of the chosen one.

\end{itemize}

During the backward pass, given $\partial L / \partial \mathbf{Y}$ and $\partial \phi (\bar{\mathbf{Y}})/{\partial \bar{\mathbf{Y}}}$, using chain rules and the above results, the derivatives required in TABL can be calculated as below:
\begin{align}
\frac{\partial L}{\partial \mathbf{W}_1} &= \frac{\partial L}{\partial \mathbf{Y}} \frac{\partial \phi (\bar{\mathbf{Y}})}{\partial \bar{\mathbf{Y}}} \frac{\partial \bar{\mathbf{Y}}}{\partial \tilde{\mathbf{X}}} \frac{\partial \tilde{\mathbf{X}}}{\partial \bar{\mathbf{X}}} \frac{\partial \bar{\mathbf{X}}}{\partial \mathbf{W}_1 } \\
\frac{\partial L}{\partial \mathbf{W}} &= \frac{\partial L}{\partial \mathbf{Y}} \frac{\partial \phi (\bar{\mathbf{Y}})}{\partial \bar{\mathbf{Y}}} \frac{\partial \bar{\mathbf{Y}}}{\partial \tilde{\mathbf{X}}} \frac{\partial \tilde{\mathbf{X}}}{\partial \mathbf{A}} \frac{\partial \mathbf{A}}{\partial \mathbf{E}} \frac{\partial \mathbf{E}}{\partial \mathbf{W}}\\
\frac{\partial L}{\partial \lambda} &= \frac{\partial L}{\partial \mathbf{Y}} \frac{\partial \phi (\bar{\mathbf{Y}})}{\partial \bar{\mathbf{Y}}} \frac{\partial \bar{\mathbf{Y}}}{\partial \tilde{\mathbf{X}}} \frac{\partial \tilde{\mathbf{X}}}{\partial \lambda} \\
\frac{\partial L}{\partial \mathbf{W}_2} &= \frac{\partial L}{\partial \mathbf{Y}} \frac{\partial \mathbf{Y}}{\partial \mathbf{W}_2} \\
\frac{\partial L}{\partial \mathbf{B}} &= \frac{\partial L}{\partial \mathbf{Y}} \frac{\partial \mathbf{Y}}{\partial \mathbf{B}}
\end{align}

\section{Complexity of Attention-based RNN}\label{Ap2}
The attention-based sequence to sequence learning proposed in \cite{bahdanau2014neural} comprises of the following modules:

\begin{itemize}
\item Encoder
\begin{align}
\mathbf{z}_i^e &= \sigma \big( \mathbf{W}_z^e \mathbf{x}_i + \mathbf{U}_z^e \mathbf{h}_{i-1}^e + \mathbf{b}_z^e \big) \label{eqA1} \\
\mathbf{r}_i^e &= \sigma \big( \mathbf{W}_r^e \mathbf{x}_i + \mathbf{U}_r^e \mathbf{h}_{i-1}^e + \mathbf{b}_r^e \big) \label{eqA2} \\
\tilde{\mathbf{h}}_i^e &= \tanh \big( \mathbf{W}^e \mathbf{x}_i + \mathbf{U}^e (\mathbf{r}_i^e \odot \mathbf{h}_{i-1}^e) + \mathbf{b}^e \big) \label{eqA3} \\
\mathbf{h}_i^e &= (1- \mathbf{z}_i^e ) \odot \mathbf{h}_{i-1}^e +\mathbf{z}_i^e \odot \tilde{\mathbf{h}}_i^e \label{eqA4}\\
\end{align}

\item Memory
\begin{align}
e_{ij} &=\mathbf{v}_{\alpha}^{T} \tanh (\mathbf{W}_{\alpha}\mathbf{h}_{j-1}^d+\mathbf{U}_{\alpha}\mathbf{h}_i^e) \label{eqA5}\\
\alpha_{ij} &=\frac{\exp (e_{ij})}{\sum_{k=1}^{T} \exp (e_{kj})} \label{eqA6}\\
\mathbf{c}_{j} &= \sum_{i=1}^{T} \alpha_{ij} \mathbf{h}_i^e  \label{eqA7}
\end{align}

\item Decoder
\begin{align}
\mathbf{z}_j^d &= \sigma \big( \mathbf{w}_z^d y_{j-1} + \mathbf{U}_z^d \mathbf{h}_{j-1}^d + \mathbf{C}_z \mathbf{c}_j + \mathbf{b}_z^d \big) \label{eqA8}\\
\mathbf{r}_j^d &= \sigma \big( \mathbf{w}_r^d y_{j-1} + \mathbf{U}_r^d \mathbf{h}_{j-1}^d + \mathbf{C}_r \mathbf{c}_j + \mathbf{b}_r^d \big) \label{eqA9}\\
\tilde{\mathbf{h}}_j^d &= \tanh \big( \mathbf{w}^d y_{j-1} + \mathbf{U}^d (\mathbf{r}_j^d \odot \mathbf{h}_{j-1}^d) + \mathbf{C} \mathbf{c}_j + \mathbf{b}^d \big) \label{eqA10}\\
\mathbf{h}_j^d &= (1- \mathbf{z}_j^d ) \odot \mathbf{h}_{j-1}^d +\mathbf{z}_j^d \odot \tilde{\mathbf{h}}_j^d \label{eqA11}\\
y_j &= \mathbf{w}_{out}^{T} \mathbf{h}_j^d + b_{out} \label{eqA12}
\end{align}
\end{itemize}

where $i=1,\dots,T$ and $j=1,\dots, T$ denote the index in input and output sequence respectively, which we assume having equal length. In order to generate sequence of prediction rather than probability of a word in a dictionary, we use Eq. (\ref{eqA12}) similar to \cite{cinar2017position}. To simplify the estimation, let the number of hidden units in the encoder, memory and decoder module equal to $D'$, i.e. $\mathbf{h}_i^e, \mathbf{h}_j^d, \mathbf{v}_{\alpha} \in \mathbb{R}^{D'}$, and the output $y_j$ is a scalar.

The encoder module estimates the following parameters: $\mathbf{W}^e, \mathbf{W}_r^e, \mathbf{W}_z^e \in \mathbb{R}^{D' \times D}$, $\mathbf{U}^e, \mathbf{U}_r^e, \mathbf{U}_z^e \in \mathbb{R}^{D' \times D'}$, $\mathbf{b}^e, \mathbf{b}_r^e, \mathbf{b}_z^e \in \mathbb{R}^{D'}$, which result in $O(3D'D+3D'^2 + 3D')$ memory and $O(T(3D'D+ 3D'^2+8D')$ computation.

The memory module estimates the following parameters: $\mathbf{v}_{\alpha} \in \mathbb{R}^{D'}$, $\mathbf{W}_{\alpha}, \mathbf{U}_{\alpha} \in \mathbb{R}^{D' \times D'}$, which cost $O(2D'^2+D')$ memory and $O(2D'^2T+4T^2D'+T^2)$ computation.

The decoder module estimates the following parameters: $\mathbf{U}_r^d, \mathbf{U}_z^d, \mathbf{U}^d, \mathbf{C}_z, \mathbf{C}_r, \mathbf{C} \in \mathbb{R}^{D' \times D'}$, $\mathbf{w}_r^d, \mathbf{w}_z^d, \mathbf{w}^d, \mathbf{b}_z^d, \mathbf{b}_r^d, \mathbf{b}^d, \mathbf{w}_{out} \in \mathbb{R}^{D'}$, which result in $O(6D'^2+7D')$ memory and $O(T(12D'+ 6D'^2))$.

In total, the attention model requires $O(3D'D+11D'^2+11D')$ memory and $O(11TD'^2+20TD'+4T^2D'+3TD'D+T^2)$ computation.

\bibliography{reference}
\bibliographystyle{ieeetr}

\end{document}